\documentclass[twocolumn,showpacs,preprintnumbers,amsmath,amssymb,prb]{revtex4-1}
\usepackage{graphicx}
\usepackage{dcolumn}
\usepackage{bm}
\usepackage[tight]{subfigure}
\usepackage{amsmath}
\usepackage{verbatim}
\usepackage{color}

\begin{document}
\title{Interaction effects in electric transport through self-assembled molecular monolayers}
\author{Martin Leijnse}
\affiliation{
      Center for Quantum Devices, Niels Bohr Institute,
      University of Copenhagen,
      2100~Copenhagen \O, Denmark 
}
\begin{abstract}
We theoretically investigate the effect of inter-molecular Coulomb interactions on transport through self-assembled 
molecular monolayers (or other devices based on a large number of nanoscale conductors connected in parallel).
Due to the interactions, the current through different molecules become correlated, resulting in distinct features in
the nonlinear current-voltage characteristics, as we show by deriving and solving a type of modified master
equation, suitable for describing transport through an infinite number of interacting conductors. 
Furthermore, if some of the molecules fail to bond to both electrodes, charge 
traps can be induced at high voltages and block transport through neighboring molecules, resulting in negative differential 
resistance.
\end{abstract}
\maketitle
\section{Introduction}
When $N$ macroscopic resistors, each with a conductance $G$, are connected in parallel, the total 
conductance is given by $G_T = N G$.
However, this is no longer true if the currents flowing through the individual resistors are correlated, i.e., if the current 
through one resistor is affected by the current through its neighbors, e.g., due to Coulomb interaction between charge carriers. 
Such correlation effects can be expected to be important in nanoscale systems, where the conductors are very close to each other 
and where, because of the quantized nature of the electron charge, current flow can be associated with significant fluctuations of 
the charge on the resistors.

One interesting example is molecular electronic devices based on a self-assembled molecular monolayer, sandwiched between 
metallic electrodes~\cite{Reed_book, Salomon03, Akkerman06}. Compared with single-molecule junctions, monolayer devices offer 
better reproducibility and stability, and the larger currents are easier to measure. They have been used to investigate a number
of interesting molecular transport effects, such as negative differential resistance (NDR)~\cite{Chen99}, switching~\cite{Donhauser01},
and spin-selective tunneling~\cite{Ray99, Gohler11}.
Although the experiments are usually interpreted within a single-molecule picture, direct experimental comparisons 
between monolayer and single-molecule junctions~\cite{Selzer05} have found large differences, i.e., $G_T \neq N G$. 
Such differences have been discussed in 
terms of static changes to the local molecular environment in a monolayer device, e.g., due to re-arrangement of molecular or 
surface charges, or interactions between constant molecular dipole moments~\cite{Naaman10, Heimel10, Egger12_mol}.

In this work, we investigate the \emph{dynamic} transport effect resulting from Coulomb interactions between charges being transported 
through neighboring molecules in a monolayer. The inter-molecular Coulomb interactions not only lower the conductance ($G_T < NG $), 
but qualitatively change the nonlinear current-voltage characteristics of the device, as we show by deriving a type of modified master
equation for the nonequilibrium current, as well as the voltage-dependent charging of the monolayer. 
If the source and drain tunnel couplings differ, the inter-molecular Coulomb interactions give rise to a voltage-asymmetric 
current-voltage characteristic, $I(V) \neq -I(-V)$, the shape of which provides an experimental fingerprint of the interactions.
Furthermore, if some molecules form bonds only with one electrode, we show that 
NDR can occur, since charge traps are formed within the layer and, through the inter-molecular 
Coulomb interaction, block transport through neighboring molecules.
Knowledge of the generic transport signatures of inter-molecular Coulomb interactions should be 
very helpful when interpreting data from molecular monolayer devices and trying to separate the genuinely single-molecule transport 
effects.
We use terminology like "molecule" and "monolayer", but the results are of much more general 
importance, and apply, for example, to transport through many quantum dots connected in parallel, carbon nanotubes bundled 
together in a rope, or arrays of nanoparticles, just to give a few examples.
The new type of master equation we develop, which describes the correlated non-equilibrium charge distribution among interacting
conductors, should form a useful starting point also for studies of a wider range of interacting mesoscopic systems.

There has been a significant amount of prior work investigating the transport effects of an inter-molecular tunnel coupling, 
see e.g., Refs.~\onlinecite{Yaliraki98, Magoga99, Reuter11b, Reuter11}. However, in contrast to Coulomb interactions, 
tunnel couplings decay exponentially with the molecular separation. 
Several works have also studied transport through two parallel-coupled quantum dots~\cite{Meden06, Chi05, Eliasen10}, 
including the effects of inter-dot Coulomb interactions and/or interference between different tunneling paths,
and Refs.~\onlinecite{Kiesslich02a, Kiesslich02b} investigated the effects of inter-dot Coulomb interactions in
finite arrays of self-assembled quantum dots. 

\section{Basic physical picture and model} 
\begin{figure}[t!]
  \includegraphics[height=0.66\linewidth]{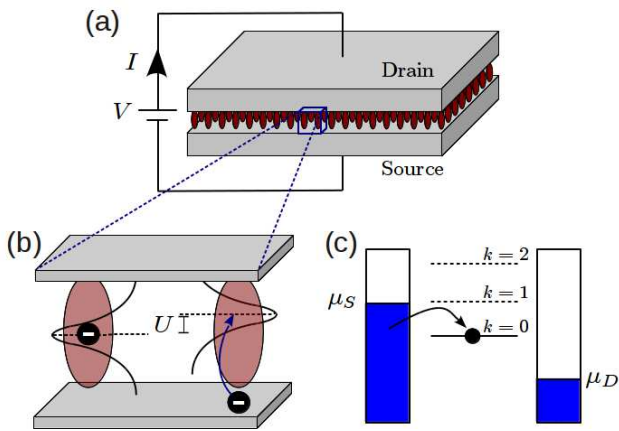}
  \caption{
    \label{fig:1}
    (Color online). (a) Sketch of the molecular monolayer device. 
    (b) Zoom in on two of the molecules in the upper panel, together with an energy-space representation of the molecular 
    transmission resonances. Due to the intermolecular Coulomb interaction, the resonance seen by an electron tunneling into a
    molecule is shifted by $U$ if one of its neighbors is charged.
    (c) Level diagram showing tunneling into the LUMO of a single molecule. The LUMO is shifted by $k U$ when 
    $k$ neighboring molecules are occupied (higher-lying dashed lines).}
\end{figure}

Figure~\ref{fig:1} shows a sketch of a molecular monolayer between metallic source and drain electrodes.
In Fig.~\ref{fig:1}(b), we focus on two molecules within the layer and illustrate the basic transport effect 
of inter-molecular Coulomb interactions, and indicate the molecular transmission resonances (the vertical direction therefore
indicates a direction both in real space and in energy space). 
When an electron tunnels into a molecule, it experiences the electric field from charges localized on neighboring molecules,
which leads to a shift of the effective transport resonance. Therefore, when the bias voltage, $V$, is just barely large enough 
to allow resonant transport through a molecular orbital (the HOMO or LUMO is inside the bias window), transport can still be 
suppressed if neighboring molecules are charged. In the limit of weak tunnel coupling, where transport is dominated by 
single-electron tunneling, and at low temperatures, we therefore expect a single molecular orbital to give rise to multiple 
transport resonances: 
One when the bias is large enough to allow tunneling into the LUMO or out of the HOMO, and additional ones when 
this becomes possible also when $k = 1, 2, \hdots$ neighboring molecules are charged, see also Fig.~\ref{fig:1}(c).
This is in contrast to the result without inter-molecular interactions, where only a single resonance occurs.
Below we calculate the detailed current-voltage curves for different interaction strengths. 
 
The distance between molecules in a monolayer depends on the molecular species and contact materials, but is typically $d \sim 1$~nm in 
a dense monolayer. A rough estimate gives the interaction between 
electrons on neighboring molecules as $U \sim e^2 / (4 \pi d \epsilon_0 \epsilon_r) \sim 1\;\mathrm{eV}/\epsilon_r$.
The relative permittivity, $\epsilon_r$, within the monolayer may be rather large and the interaction strength further
decreased by screening from conduction electrons in the electrodes~\cite{Kaasbjerg08}.
However, a $U$ of the same order of magnitude as the thermal energy at room temperature still seems reasonable. 
In other mesoscopic systems, $U$ is likely much smaller ($U = 0.15 - 0.4$~meV between strands of 
nanotubes within a nanotube rope was found in recent experiments~\cite{Goss11, Goss12}), but can still dominate the 
transport physics at low temperatures.

To focus on the interaction effects, we study the simplest possible molecular model, including only a single spinless orbital (chosen as 
the LUMO for definiteness). This model describes the relevant physics when the inter-molecular Coulomb interaction is much smaller 
than the local (intra-molecular) Coulomb interaction and the molecular level spacing, such that these other energy scales are irrelevant 
at low bias voltages. The Hamiltonian of the monolayer device is 
$H = \sum_i ( H^{i}_\mathrm{ML} + H_T^i ) + H_\mathrm{res}$, where 
\begin{eqnarray}
  \label{eq:HML}
	H_\mathrm{ML}^{i}	&=& \epsilon_i n_i + \frac{1}{2} \sum_j U_{i j} n_i n_j, \\
  \label{eq:Hres}
	H_\mathrm{res}	 	&=& \sum_{p r} \epsilon_{p r} n_{p r}, \\
  \label{eq:HT}
	H_T^i 			&=& \sum_{p r} t_{r i} c_{p r}^\dagger d_i + h.c. .
\end{eqnarray}
Here, $\epsilon_i$ is the energy of the LUMO of molecule $i$, which has occupation $n_i = d_i^\dagger d_i$, and
$U_{i j}$ is the Coulomb charging energy between electrons on dots $i$ and $j$ (in all calculations we include only 
nearest neighbor interactions, although the theoretical framework developed below can straightforwardly be extended to include 
more long-ranged interactions). 
The source ($r=S$) and drain ($r = D$) electrodes are described by $H_\mathrm{res}$, where $n_{p r} = c_{p r}^\dagger c_{p r}$ is
the number operator for electrons in state $p$. The electrons in the electrodes are as usual for good metals modelled as being non-interacting.
$ H_T^i $ describes tunneling between electrode $r = S,D$ and molecule $i$, which takes place with amplitude $t_{r i}$. We assume both the 
tunnel amplitude and the electrode densities of states, $\rho_r$, to be energy-independent. In this case, the tunnel rates, 
$\Gamma_{r i} = 2 \pi \rho_r |t_{ri}|^2 / \hbar$, which set the inverse time-scale for single-electron tunneling, are also energy-independent. 

\section{Master equation for transport through a homogeneous layer}
We consider first a homogeneous monolayer, i.e., $U_{ij} = U$, $\Gamma_{ri}$ = $\Gamma_r$, and $\epsilon_i = \epsilon$.
We also focus on the regime where transport is dominated 
by single-electron tunneling, which is the case in the weak tunnel coupling regime, $\hbar \Gamma < k_B T$, where $T$ is the temperature.
However, since we will focus primarily on resonant (or close to resonant) transport, corrections from higher order tunnel processes, such 
as elastic and inelastic cotunneling, are expected to only change the results quantitatively even in the regime $\hbar \Gamma \sim k_B T$.
We arbitrarily pick one molecule in the monolayer and denote it by M1. Our aim is now to calculate the stationary current 
flowing through the LUMO of M1. In the single-electron tunneling regime, the current is carried by processes in which a single 
electron tunnels either from one electrode into M1, or from M1 into one electrode, with rates 
proportional to the probability that M1 is empty, $P^{(0)}$, or full, $P^{(1)}$, respectively. However, because of $U$, the rates also depend on 
the occupation of the $N$ nearest neighbors of M1, which we denote M2. We therefore consider probability distributions 
of the form $P^{(m)}_k$: The probability that M1 is empty ($m=0$) or occupied ($m = 1$), while $k$ of its neighbors are occupied.
The net current from reservoir $r$ into M1 is then
\begin{eqnarray}
\label{current}
	I_r	&=& -e \sum_{k=0}^{N} \left( W_r^+(k U) P^{(0)}_k - W_r^-(k U) P^{(1)}_k\right),
\end{eqnarray}
where $W_r^{\pm}(E) = \Gamma_r f^{\pm}((E + \epsilon - \mu_r)/k_BT)$, $f^{+}(x)$ is the fermi function,
$f^{-}(x) = 1 - f^{+}(x)$, and $\mu_r = \pm e V/2$ for $r=S/D$ is the chemical potential of electrode $r$.
We assume that the bias drop occurs symmetrically at the two tunnel barriers, such that $\epsilon$ is independent 
of $V$. The total current is $I = N_M I_D = - N_M I_S$, where $N_M$ is the total number of molecules in the monolayer.

The remaining problem is to find a master equation to determine $P^{(m)}_k$ under non-equilibrium conditions,
i.e., to find the voltage-induced charging of the interacting monolayer. Just as in a normal master equation, the time
derivative of an occupation probability is given by the sum of all tunnel processes enhancing that occupation, 
minus all tunnel processes decreasing that occupation, each process being weighted by the occupation probability 
of the corresponding initial state. We obtain
\begin{eqnarray}\label{ME1}
	\dot{P}_k^{(1)} 	&=& W^+ (k U) P_k^{(0)} - W^- (k U) P_k^{(1)}\nonumber \\
  				&-& N \sum^N_{k' = 1} \left[ W^- (k' U) P_{k, k'}^{(1, 1)} + W^+ (k' U) P_{k,k'}^{(1, 0)} \right]  P_k^{(1)} \nonumber \\
  				&+& N \sum^N_{k' = 1} \left[ W^- (k' U) P_{k + 1}^{(1)} P_{k + 1, k'}^{(1, 1)} \right. \nonumber \\
				&+& \left. W^+ (k' U) P_{k - 1}^{(1)} P_{k - 1, k'}^{(1, 0)} \right],
\end{eqnarray}
\begin{eqnarray}\label{ME0}
  \dot{P}_k^{(0)} 		&=& -W^+ (k U) P_k^{(0)}  + W^- (k U)  P_k^{(1)}  \nonumber \\
  				&-& N \sum^{N - 1}_{k' = 0} \left[ W^- (k' U) P_{k, k'}^{(0, 1)} +  W^+ (k' U) P_{k, k'}^{(0, 0)} \right] P_k^{(0)} \nonumber \\
  				&+& N \sum^{N - 1}_{k' = 0} \left[ W^-(k' U) P_{k + 1}^{(0)} P_{k + 1, k'}^{(0, 1)} \right. \nonumber \\
				&+& \left. W^+ (k' U) P_{k - 1}^{(0)} P_{k - 1, k'}^{(0, 0)} \right],
\end{eqnarray}
where $W^{\pm} (E) = \sum_r W_r^{\pm} (E)$. The first line in Eqs.~(\ref{ME1}) and~(\ref{ME0}) is just like in a
standard master equation, describing the increase (decrease) in occupation of M1 due to tunneling into (out of) it. 
The second (third and forth) lines describe the decrease (increase) in $P_k^{(m)}$ due to tunneling into or out of one of M2 
(nearest neighbors of M1), which changes $k$. Here, $P_{k, k'}^{(1, 1)}$, for example, denotes the
\emph{conditional} probability that an M2 is occupied and has $k'$ occupied neighbors, \emph{given that} M1 is occupied with $k$
occupied nearest neighbors. Note that $k'$ runs from 1 to $N$ in the equation for $P_k^{(1)}$, but from 0 to $N - 1$ 
in the equation for $P_k^{(0)}$. The reason is that if M1 is occupied, M2 has to have at least one occupied nearest
neighbor, and if M1 is unoccupied, M2 can at the most have $N-1$ occupied neighbors.

In the steady-state limit, which is assumed in all results presented here, we can set 
all time derivatives to zero. 
Equations~(\ref{ME1}) and~(\ref{ME0}) have to be supplemented with a condition expressing probability normalization:
\begin{eqnarray}\label{normalize}
	\sum_k \left( P_k^{(0)} + P_k^{(1)} \right) = 1.
\end{eqnarray}
The average charge on M1 is given by $\langle q \rangle = -e \sum_k P^{(1)}_k$.

Clearly we need an additional master equation to solve for $P^{(m, n)}_{k, k'}$, which can be derived as a standard master 
equation by considering all ingoing and outgoing processes. 
Since this equation is rather lengthy, it is given in Appendix~\ref{sec:appA}.
The master equation for $P^{(m, n)}_{k, k'}$ in turn involves higher order conditional probabilities, $P^{(m, n, l)}_{k, k', k''}$, 
and so on. 
To close the system of equations involving ever higher orders of conditional probabilities we need to invoke some kind of approximation. 
Appendix~\ref{sec:appB} discusses an advanced type of 
mean-field approximation, which essentially treats interactions between M1 and M2 molecules exactly, while interactions between 
M2 and M3 molecules are treated in a mean-field manner. This produces reliable results only for rather small interaction strengths, 
$U \lesssim k_B T$. We want to investigate also the regime $U > k_B T$ and in all results presented below we therefore 
apply a more advanced approximation scheme. 
We close the hierarchy of equations by neglecting explicit three-charge 
correlations (replacing $P^{(m, n, l)}_{k, k', k''} \rightarrow P^{(n, l)}_{k', k''}$). One is then left with solving a nonlinear 
master equation for $P^{(m, n)}_{k, k'}$, which have to be inserted into the equation for $P^{(m)}_{k}$. This approximation 
and the explicit form of the resulting nonlinear master equation is discussed in detail in Appendix~\ref{sec:appC}.
Appendix~\ref{sec:appE} compares the different approximation schemes.  

\section{Current--voltage characteristics for homogeneous layers}
\begin{figure}[t!]
  \includegraphics[height=0.7\linewidth]{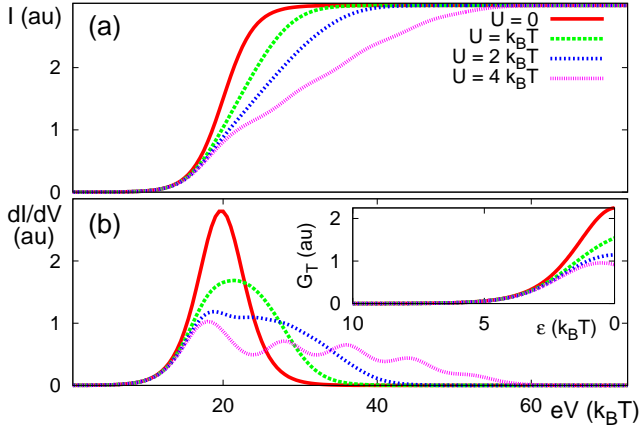}
  \caption{
    \label{fig:2}
    (Color online). $I(V)$, (a), and $dI(V)/dV$, (b), for a homogeneous monolayer with $\epsilon = 10 k_B T$ and 
    increasing strength of the inter-molecular Coulomb interaction, $U = 0, \hdots ,4 k_B T$.
    We have assumed a square lattice configuration, where each molecule has $N = 4$ 
    nearest neighbors. $I$ and $dI/dV$ are given in arbitrary 
    units (au) since $I \propto \Gamma$ in the single-electron tunneling regime.
    The inset in (b) shows the linear conductance, $G_T$, as a function of the position of the LUMO relative to the 
    Fermi energy.}
\end{figure}

We now find the current from Eq.~(\ref{current}), with probability distributions calculated from Eqs.~(\ref{ME1}) and~(\ref{ME0})
and conditional probabilities found within the approximation scheme in Appendix~\ref{sec:appC} from Eqs.~(\ref{CMEcorr11})--(\ref{CMEcorr01}).
Figure~\ref{fig:2} shows the results for increasing interaction strengths, $U = 0, \hdots, 4 k_B T$. 
When $U = 0$, $I(V)$ shows a thermally broadened step when the LUMO 
enters into the bias window, which acquires some additional broadening for $U \lesssim  k_B T$. When $U > k_B T$, several current steps 
(multiple sidepeaks in $dI/dV$) can be discerned. They are separated in voltage by $U$ and correspond to the condition
that charging M1 becomes possible, also when $k=1, 2, \hdots, N$ of M2 are charged.
There are only $N$ satellite conductance peaks, since the maximum 
interaction cost of adding one electron is $NU$, and the peaks are always equidistant. This could help to 
experimentally distinguish sidepeaks originating from inter-molecular Coulomb interaction from similar features related to higher lying 
orbitals or vibronic excitations~\cite{Park00, Tao06rev, Galperin07}.
For all values of $U$, the current saturates at the same value, corresponding to a fully conducting monolayer. 
The inset of Fig.~\ref{fig:2}(b) shows the linear conductance as a function of the position of the LUMO relative to the 
electrode Fermi energy. Unless the LUMO is close to resonance, the charging of the monolayer is rather small and the
inter-molecular Coulomb interactions are relatively unimportant.

\begin{figure}[t!]
  \includegraphics[height=0.7\linewidth]{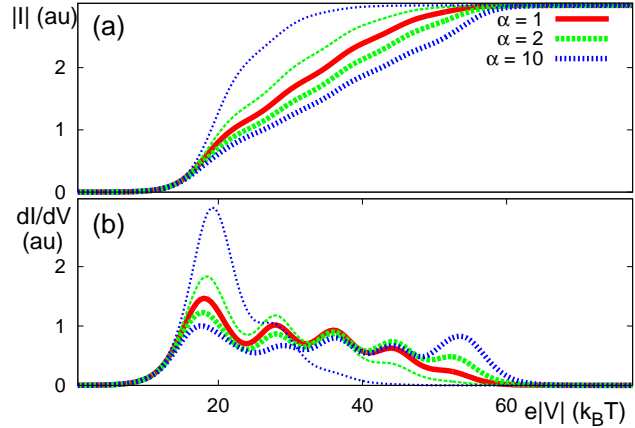}
  \caption{
    \label{fig:3}
    (Color online). $|I|$, (a), and $dI/dV$, (b), as a function of $V$ (thick lines) or $-V$ (thin lines), for $\epsilon = 10 k_B T$,
    $U = 4 k_B T$, and $N = 4$. We vary the ratio between the source and drain tunnel couplings, $\alpha = \Gamma_S / \Gamma_D$, while 
    keeping $\bar{\Gamma} = \Gamma_S \Gamma_D / (\Gamma_S + \Gamma_D)$ fixed, which fixes the value of the current plateau at large $V$.}
\end{figure}

Figure~\ref{fig:3} shows $I (V)$ and $dI(V) / dV$ for a monolayer with a larger coupling to the source than 
the drain, $\Gamma_S > \Gamma_D$. For $U \neq 0$, this introduces an asymmetry into the current voltage characteristics, 
$I (V) \neq -I(-V)$.
The reason is that there is now a larger charging of the monolayer for $V > 0$ compared with $V < 0$ since, for positive bias, electrons 
can easily tunnel into the monolayer (from the source), but cannot easily escape again (to the drain), while the situation is 
reversed for negative bias. A larger charging increases the effects of interactions and causes the $I(V)$ curve 
to become flatter and $dI/dV$ to show more pronounced sidepeaks.
For $U = 0$, asymmetric tunnel couplings do not introduce any asymmetry into the $I(V)$ curve. There are several other possible 
reasons for an asymmetric $I(V)$. However, the rather special form seen in Fig.~\ref{fig:3}, with the same plateau height for positive and 
negative bias but with clear multiple peaks only for one bias polarity, nonetheless provides a fingerprint of the inter-molecular 
Coulomb interaction.

\section{NDR in inhomogeneous layers}
Finally, we investigate the effects of disorder within the molecular monolayer. Here, it is no longer possible to use our
master equation approach, which assumes a homogeneous monolayer. Instead, we take a finite number of molecules, which may all
have different properties, and diagonalize the corresponding many-body Hamiltonian exactly. The many-body eigenstates can then 
be used in a standard master equation approach~\cite{Bruus04book} to calculate the current. 
The details are given in Appendix~\ref{sec:appD}.
Due to the rapidly growing size of the Hilbert space, only a rather small number of molecules can be included. 
However, as is shown in Appendix~\ref{sec:appE} for the case of a homogeneous monolayer, using
$3\times3$ molecules and periodic boundary conditions gives excellent agreement with the infinite monolayer master 
equation.

\begin{figure}[t!]
  \includegraphics[height=0.37\linewidth]{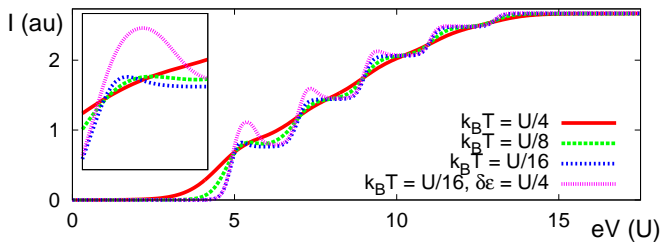}
  \caption{\label{fig:4} (Color online). $I (V)$ for $\epsilon = 5 U / 2$, $N = 4$, and
    successively lowered $T$. 
    One molecule (out of nine) has $\Gamma_D = \Gamma_S / 100$. 
    In the magenta curve, we have furthermore assumed the LUMO of the asymmetrically coupled 
    molecule to be $\delta \epsilon = U/4$ higher in energy compared with the others.
    The inset zooms in on the first current step.}
\end{figure}

Figure~\ref{fig:4} shows the result of these calculations, where we have assumed one molecule (out of nine) to be only weakly 
coupled to the drain electrode, $\Gamma_D = \Gamma_S / 100$. 
The asymmetrically tunnel coupled molecule now acts as a charge trap, since for $V > 2\epsilon$ it can be filled with an electron  
tunneling in from the source, which is then prevented from escaping to the drain. The inter-molecular Coulomb interaction means that
the occupation of the charge trap prevents transport through neighboring molecules, unless the bias voltage is increased further.
As the temperature is reduced, this leads to a weak NDR at the first current plateau. 
A much stronger NDR effect, which is repeated on every current plateau, is obtained if the LUMO of the asymmetrically 
coupled molecule lies at a slightly higher energy compared 
with the other molecules (see magenta curve in Fig.~\ref{fig:4}). This is a realistic scenario since the asymmetrically coupled 
molecule experiences less screening from conduction electrons in the electrodes and, in general, a different chemical and electrostatic 
environment. The reason for the larger NDR is that, due to the misalignment, a large current is first allowed to flow through the 
symmetrically coupled molecules, before the charge trap becomes filled at higher voltages.
For the configuration in Fig.~\ref{fig:4}, with a molecule weakly coupled to the drain, the $I(V)$ curve will be asymmetric, with no 
NDR for negative bias. However, in a large monolayer, if approximately the same number of molecules are weakly coupled to the source and 
to the drain, the voltage symmetry is approximately restored with NDR for both bias polarities. 

\section{Conclusions}
We have studied electric transport through a large number of mesoscopic conductors contacted in parallel, and
shown that Coulomb interactions between charge carriers on 
neighboring conductors give rise to distinct features in the nonlinear current-voltage characteristics. 
Knowledge of the interaction effects are relevant when designing nanoscale electric devices, and
our calculations allow the interaction-induced features to be identified and isolated from the single-device 
properties, which is essential when using electric transport as a spectroscopic tool. Furthermore, interactions
can give rise to negative differential resistance if some of the conductors are significantly coupled to only one 
electrode, which may be useful for a range of applications.
There has been much recent interest in molecular thermoelectric devices~\cite{Reddy07, Dubi11} because of the sharp nature 
of the transport resonances~\cite{Murphy08}. The results of our study, showing an effective interaction-induced broadening of 
the transport resonances, is likely to have a large impact on the thermoelectric efficiency.
Finally, the general theoretical framework 
should form a useful basis for future studies of transport through interacting mesoscopic conductors.
The use of (conditional) probability distributions, which can be obtained from a master equation, 
to describe interacting systems could be useful also for studies of other correlated systems out of equilibrium.

\section*{Acknowledgements}
We thank Karsten Flensberg and Gemma Solomon for discussions and for useful comments on the manuscript.
The research leading to these results was carried out partly in the Danish-Chinese Centre for Molecular 
Nano-Electronics, supported by the Danish National Research Foundation, and 
has received funding from the European Union Seventh Framework Programme
(FP7/2007-2013) under agreement no 270369 (“ELFOS”).
The Center for Quantum Devices is funded by the Danish National Research Foundation.

\appendix
\section{Master equations for conditional probabilities}\label{sec:appA}

The master equation for the conditional probabilities, $P_{k, k'}^{(m, n)}$, are
\begin{widetext}
\begin{eqnarray}\label{CME11}
  \dot{P}_{k, k'}^{(1, 1)} 	&=&  W^+ (k' U) P_{k, k'}^{(1, 0)} - W^- (k' U) P_{k, k'}^{(1, 1)}  \nonumber \\
  				&-& N_- \sum^N_{k'' = 1} \left[  W^- (k'' U) P_{k, k', k''}^{(1, 1, 1)}
				    + W^+ (k'' U) P_{k, k', k''}^{(1, 1, 0)} \right] P_{k, k'}^{(1, 1)} \nonumber\\
  				&+& N_- \sum^N_{k'' = 1} \left[ W^- (k'' U) P_{k, k' + 1}^{(1, 1)} P_{k, k' + 1,k''}^{(1, 1, 1)}
				    + W^+ (k'' U) P_{k, k' - 1}^{(1, 1)} P_{k, k' - 1,k''}^{(1, 1, 0)} \right],
\end{eqnarray}
\begin{eqnarray}\label{CME00}
  \dot{P}_{k, k'}^{(0, 0)} 	&=& - W^+ (k' U) P_{k, k'}^{(0, 0)}  + W^- (k' U) P_{k, k'}^{(0, 1)} \nonumber \\
  				&-& N_- \sum^{N - 1}_{k'' = 0} \left[ W^- (k'' U) P_{k, k',k''}^{(0, 0, 1)} + 
				    W^+ (k'' U) P_{k, k', k''}^{(0, 0, 0)} \right] P_{k, k'}^{(0, 0)} \nonumber \\
  				&+& N_- \sum^{N - 1}_{k'' = 0} \left[ W^- (k'' U) P_{k, k' + 1}^{(0, 0)} P_{k, k' + 1, k''}^{(0, 0, 1)} + 
				    W^+ (k'' U) P_{k, k' - 1}^{(0, 0)} P_{k, k' - 1, k''}^{(0, 0, 0)} \right],
\end{eqnarray}
\begin{eqnarray}\label{CME10}
  \dot{P}_{k, k'}^{(1, 0)} 	&=& -W^+ (k' U) P_{k, k'}^{(1, 0)} + W^- (k' U) P_{k, k'}^{(1, 1)} \nonumber \\
  				&-& N_- \sum^{N - 1}_{k'' = 0} \left[ W^- (k'' U) P_{k, k', k''}^{(1, 0, 1)} 
				    + W^+ (k'' U)  P_{k, k', k''}^{(1, 0, 0)} \right] P_{k, k'}^{(1, 0)} \nonumber \\
 	 			&+& N_- \sum^{N - 1}_{k'' = 0} \left[ W^- (k'' U) P_{k, k' + 1}^{(1, 0)} P_{k, k' + 1, k''}^{(1, 0, 1)}
				    + W^+ (k'' U) P_{k, k' - 1}^{(1, 0)} P_{k, k' - 1, k''}^{(1, 0, 0)} \right],
\end{eqnarray}

\begin{eqnarray}\label{CME01}
  \dot{P}_{k, k'}^{(0, 1)} 	&=& W^+ (k' U) P_{k, k'}^{(0, 0)}  - W^- (k' U) P_{k, k'}^{(0, 1)} \nonumber \\
  				&-& N_- \sum^N_{k'' = 1} \left[ W^- (k'' U) P_{k, k', k''}^{(0, 1, 1)} + 
				    W^+ (k'' U) P_{k, k', k''}^{(0, 1, 0)} \right] P_{k, k'}^{(0, 1)} \nonumber \\
  				&+& N_- \sum^N_{k'' = 1} \left[ W^- (k'' U) P_{k, k' + 1}^{(0, 1)} P_{k, k' + 1, k''}^{(0, 1, 1)}+ 
				    W^+ (k'' U) P_{k, k' - 1}^{(0, 1)} P_{k, k' - 1, k''}^{(0, 1, 0)} \right],
\end{eqnarray}
\end{widetext}
where $N_- = N - 1$. Here, $P_{k, k', k''}^{(1, 1, l)}$, for example, denotes the conditional probability that a 
next-nearest neighbor molecule (M3) is occupied ($l=1$) or empty ($l=0$), with $k''$ occupied nearest neighbors, given that M1 is occupied with $k$ 
occupied nearest neighbors and M2 is occupied with $k'$ occupied nearest neighbors. 

Just as the normal occupation probabilities, the conditional probabilities must be normalized. This gives rise to the additional equations:
\begin{eqnarray}\label{Cnormalize0}
  1 	&=& \sum^{N - 1}_{k' = 0} \left( P_{k, k'}^{(0, 0)} + P_{k, k'}^{(0, 1)} \right),\\
\label{Cnormalize1}
  1 	&=& \sum^N_{k' = 1} \left( P_{k, k'}^{(1, 0)} + P_{k, k'}^{(1, 1)} \right).
\end{eqnarray}
In addition, we have 
\begin{eqnarray}\label{zeros}
	0 = P_{k, N}^{(0, j)} = P_{k, 0}^{(1, j)} = P_{N, k'}^{(i, 0)} = P_{0, k'}^{(i, 1)},
\end{eqnarray}
since, for example, M2 must have at least one occupied neighbor if M1 is occupied. 

There is a further possible simplification due to the homogeneous nature of the monolayer: Since all M2 molecules are equivalent, 
if M1 has $k$ occupied neighbors, each M2 should be occupied with probability $k / N$. This gives the conditions
\begin{eqnarray}\label{CEnormalize1}
  \frac{k}{N} 		&=& \sum^N_{k' = 0} P_{k, k'}^{(m, 1)}, \\
\label{CEnormalize0}
  1 - \frac{k}{N} 	&=& \sum^N_{k' = 0} P_{k, k'}^{(m, 0)}.
\end{eqnarray}
When $P_{k, k'}^{(m, n)}$ fulfill Eqs.~(\ref{CEnormalize1}) and~(\ref{CEnormalize0}), they also automatically fulfill the standard normalization 
conditions, Eqs.~(\ref{Cnormalize0}) and~(\ref{Cnormalize1}). However, Eqs.~(\ref{CEnormalize1}) and~(\ref{CEnormalize0}) give twice 
as many conditions as Eqs.~(\ref{Cnormalize0}) and~(\ref{Cnormalize1}), and should be used instead of the first line in Eqs.~(\ref{CME11})--(\ref{CME01}),
corresponding to the terms changing $n$ in $P_{k, k'}^{(m, n)}$. Thus, the occupation of M2 
should not be determined by the master equation, but is instead fixed by $k$. We therefore use the master equation only to solve for the 
$k'$-dependence of $P_{k, k'}^{(m, n)}$.

\section{Mixed mean-field approximation}\label{sec:appB}
We start with a simple approximation to close the system of master equations involving ever higher orders of conditional 
probabilities. 
This approximation is not used in any of the results in the results presented in Figs.~\ref{fig:2}--\ref{fig:4}, 
but is included here since it provides a simple 
and intuitive method, which is valid for moderate values of the interaction strength, $U \lesssim k_B T$.
We treat the interactions between M1 and M2 exactly, while the interactions between M2 and M3 are
treated within a mean-field approximation. In the master equations for $P_k^{(m)}$,
we should then replace $k' U$ by $N_- \langle n \rangle U$ in Eq.~(\ref{ME0}) (for $m=0$) and by $(1 + N_- \langle n \rangle) U$ in 
Eq.~(\ref{ME1}) (for $m=1$), where $\langle n \rangle$ is the average occupation of the $N+1$ molecules M1 and M2, given by
\begin{eqnarray}\label{occaverage}
\langle n \rangle = \frac{1}{N + 1} \sum_{k'} \left[ P^{(1)}_{k'} + N k'
   \left(P^{(1)}_{k'} + P^{(0)}_{k'} \right) \right].
\end{eqnarray}
After this approximation, $W^{\pm}$ in Eqs.~(\ref{ME1}) and~(\ref{ME0}) no longer depend on $k'$ and 
we can use Eqs.~(\ref{CEnormalize1}) and~(\ref{CEnormalize0}) to carry out the sums over the conditional probabilities 
$P_{k, k'}^{(m, n)}$. Thus, we do not need to solve Eqs.~(\ref{CME11})--(\ref{CME01}), but can directly solve 
Eqs.~(\ref{ME1}) and~(\ref{ME0}), which have now become
\begin{widetext}
\begin{eqnarray}\label{MMF1}
  \dot{P}_k^{(1)} 	&=&  W^+ (k U) P_k^{(0)} - W^- (k U) P_k^{(1)}  \nonumber \\
  			&-& N \left\{ \frac{k}{N} W^- \left[ (1 + N_- \langle n \rangle) U \right] + 
			    \left( 1 - \frac{k}{N} \right) W^+ \left[ (1 + N_- \langle n \rangle) U \right] \right\} P_k^{(1)} \nonumber \\
  			&+& N \left\{ \frac{k + 1}{N} W^- \left[(1 + N_- \langle n \rangle) U \right]  P_{k + 1}^{(1)} + 
			    \left( 1 - \frac{k - 1}{N} \right) W^+ \left[ (1 + N_- \langle n \rangle) U\right] P_{k - 1}^{(1)} \right\},
\end{eqnarray}
\begin{eqnarray}\label{MMF0}
  \dot{P}_k^{(0)} 	&=& W^- (k U) P_k^{(1)} - W^+ (k U) P_k^{(0)} \nonumber\\
  			&-& N \left[ \frac{k}{N} W^- \left( N_-\langle n \rangle U \right) + 
			    \left( 1 - \frac{k}{N} \right) W^+ \left( N_- \langle n \rangle U \right)\right] P_k^{(0)} \nonumber \\
  			&+& N \left[ \frac{k + 1}{N} W^- \left( \langle n \rangle N_- U \right) P_{k + 1}^{(0)} +
  			    \left( 1 - \frac{k - 1}{N} \right) W^+ \left( N_- \langle n \rangle U \right) P_{k - 1}^{(0)} \right].
\end{eqnarray}
\end{widetext}
Together with probability normalization, Eq.~(\ref{normalize}), this makes a closed set of equations, which, however, has to be 
solved self-consistently together with Eq.~(\ref{occaverage}) because of the dependence on $\langle n\rangle$.
The resulting occupations, $P^{(m)}_k$, can then be inserted into Eq.~(\ref{current}) to calculate the current.

Note that in Eqs.~(\ref{MMF1}) and~(\ref{MMF0}), and in Eq.~(\ref{current}) for the current, $kU$ appears explicitly in the terms 
involving tunneling into and out of M1 and, in addition, the relevant interaction for tunneling into M2 is $(1 + N_- \langle n \rangle)U$
if M1 is occupied [Eq.~(\ref{MMF1})], but $N_- \langle n \rangle U$ if M1 is unoccupied [Eq.~(\ref{MMF0})].
This reflects the fact that the M1--M2 interactions are treated exactly. 
Therefore, these master equations are capable of producing steps in the nonequilibrium monolayer charging and current, for example as a function 
of increasing bias voltage $V$, and is applicable for $U \lesssim k_B T$, see Fig.~\ref{fig:S1}. 
This is in contrast to the much simpler standard mean-field approximation, which would consist of simply 
solving only the first line in Eqs.~(\ref{MMF1}) and~(\ref{MMF0}), with $kU \rightarrow P^{(1)} U = \sum_k P^{(1)}_k U$, in which case 
the whole equation can be summed over $k$. The standard mean-field approach fails completely unless $U \ll k_B T$ (not shown in Fig.~\ref{fig:S1}).

Note also that since we treat the M1 and M2 molecules within a different level of approximation, it may happen that 
$\langle n \rangle \neq P^{(1)}$,
although we numerically always find them to be comparable, unless $U \gg k_B T$.
Similarly, if we instead of using Eq.~(\ref{current}) to calculate the current, calculated the average current through M1 and 
M2, we would obtain a slightly different result. 

\section{Truncating the correlations}\label{sec:appC}
We now introduce the approximation scheme which is used for the calculations presented in Figs. 2 and 3, 
which does not rely on a mean-field treatment at any level and is applicable also in the regime $U > k_B T$.
This is obtained by only taking conditional probabilities up to some given order explicitly into account. Here, we simply neglect 
explicit three-charge correlation terms when solving the set of master equations, which means that we 
replace $P_{k, k', k''}^{(m, n, l)} \rightarrow P_{k', k''}^{(n, l)}$ (note that since we are dealing with conditional probabilities, 
we should not factorize the expressions, i.e., we should \emph{not} include a factor $P_k^{(m)}$ on the right hand side). 
Now the equations for the conditional probabilities, Eqs.~(\ref{CME11})--(\ref{CME01}), become
\begin{widetext}
\begin{eqnarray}\label{CMEcorr11}
  \dot{P}_{k, k'}^{(1, 1)} 	&=& -N_- \sum^N_{k''= 1} \left[ W^- (k'' U) P_{k', k''}^{(1, 1)}+ 
				    W^+ (k'' U) P_{k', k''}^{(1, 0)} \right] P_{k, k'}^{(1, 1)} \nonumber \\
  				&+& N_- \sum^N_{k'' = 1} \left[ W^- (k'' U) P_{k, k' + 1}^{(1, 1)} P_{k' + 1, k''}^{(1, 1)} + 
				    W^+ (k''U) P_{k, k' - 1}^{(1, 1)} P_{k' - 1, k''}^{(1, 0)} \right],
\end{eqnarray}
\begin{eqnarray}\label{CMEcorr00}
  \dot{P}_{k, k'}^{(0, 0)} 	&=& -N_- \sum^{N - 1}_{k'' = 0} \left[ W^- (k'' U) P_{k', k''}^{(0, 1)} + 
				    W^+(k'' U) P_{k', k''}^{(0, 0)} \right] P_{k, k'}^{(0, 0)} \nonumber \\
		  		&+& N_- \sum^{N - 1}_{k'' = 0} \left[ W^- (k'' U) P_{k, k' + 1}^{(0, 0)} P_{k' + 1, k''}^{(0, 1)} + 
				    W^+ (k'' U) P_{k, k' - 1}^{(0, 0)} P_{k' - 1, k''}^{(0, 0)} \right],
\end{eqnarray}
\begin{eqnarray}\label{CMEcorr10}
\dot{P}_{k, k'}^{(1, 0)} 	&=& -N_- \sum^{N - 1}_{k'' = 0} \left[ W^- (k'' U) P_{k', k''}^{(0, 1)} + 
				    W^+(k'' U) P_{k', k''}^{(0, 0)}  \right] P_{k, k'}^{(1, 0)} \nonumber \\
  				&+& N_- \sum^{N - 1}_{k'' = 0} \left[ W^- (k'' U) P_{k, k' + 1}^{(1, 0)} P_{k' + 1, k''}^{(0, 1)} + 
				    W^+ (k'' U) P_{k, k' - 1}^{(1, 0)} P_{k' - 1, k''}^{(0, 0)} \right],
\end{eqnarray}
\begin{eqnarray}\label{CMEcorr01}
  \dot{P}_{k, k'}^{(0, 1)} 	&=& - N_- \sum^N_{k'' = 1} \left[ W^- (k'' U) P_{k', k''}^{(1, 1)} + 
				    W^+ (k'' U) P_{k', k''}^{(1, 0)} \right] P_{k, k'}^{(0, 1)} \nonumber \\
  				&+& N_- \sum^N_{k'' = 1} \left[ W^- (k'' U) P_{k, k' + 1}^{(0, 1)} P_{k' + 1, k''}^{(1, 1)} + 
				    W^+ (k''U) P_{k, k' - 1}^{(0, 1)} P_{k' - 1, k''}^{(1, 0)} \right].
\end{eqnarray}
\end{widetext}
Together with Eqs.~(\ref{CEnormalize1}) and~(\ref{CEnormalize0}), these equations can be solved for $P_{k, k'}^{(m, n)}$.
The conditional probabilities are then inserted into the master equations for the occupation probabilities, 
Eqs.~(\ref{ME1}) and~(\ref{ME0}), which can be solved for $P^{(m)}_k$, from which we then calculate the current using Eq.~(\ref{current}).

One complication is that Eqs.~(\ref{CMEcorr11})--(\ref{CMEcorr01}) are nonlinear, involving terms like
$P_{k, k'}^{(m, n)} P_{k', k''}^{(n, l)}$. Numerically we can deal with this simply by picking some initial distribution for the 
conditional probabilities which originate from the truncated higher order terms ($P_{k', k''}^{(n, l)}$), and then solve the equations 
iteratively, updating this distribution each time. Note that what remains are only equations for each distribution as a
function of $k'$, i.e., we only need to solve for the $k'$-dependence of $P_{k, k'}^{(m, n)}$. Thus, we only ever have to solve
equations of size $N \times N$, although there are $4 \times N$ such equations (one for each $m$, $n$, and $k$) which all 
have to be solved within each iteration. As demonstrated in Fig.~\ref{fig:S1}, the results are reliable for much larger values of $U/k_B T$ compared 
with the mixed mean-field approximation introduced above.

\section{Exact diagonalization of a finite monolayer}\label{sec:appD}
To study an inhomogeneous 
monolayer, e.g., involving defect sites, we rely on a standard master equation solution for a finite monolayer.
We take a finite number of molecules 
and diagonalize exactly the corresponding monolayer Hamiltonian, $\sum_i H_\mathrm{ML}^i$ in Eq.~(1).
Since the total charge in the monolayer, $N_C$, commutes with $\sum_i H_\mathrm{ML}^i$, we can label the many-body eigenstates 
with $N_C$, and, since we do not consider inter-molecular tunneling, each eigenstate can be chosen to correspond to a given
distribution of these charges over the different molecules. We thus label the eigenstates $|a N_C\rangle$, where $a$ 
labels the charge configuration. The master equation is now solved for the occupation probabilities, $p_{a N_C}$, 
of these eigenstates, which can then be used to calculate the steady-state current from electrode $r$ into the monolayer
\begin{eqnarray}\label{SME}
	\dot{p}_{a N_C} &=& \sum_{s=\pm 1} \sum_{a'}\left(K^s_{a N_C, a' (N_C + s)} p_{a' (N_C + s)} \right. \nonumber \\
		     	&-& \left. K^{-s}_{a' (N_C + s), a N_C} p_{a N_C} \right), \\
\label{Snormalize}
	1		&=& \sum_{a N_C} p_{a N_C}, \\
\label{Scurrent}
	I_r		&=& -e \sum_{a N_C} \sum_{s=\pm1} \sum_{a'} s K^{r, s}_{a' (N_C + s), a N_C} p_{a N_C},
\end{eqnarray}
where the rate matrix is given by
\begin{eqnarray}\label{Kdef}
	K^{r, s}_{a N_C, a' (N_C + s)} 	&=& \Gamma_{a N_C, a' (N_C + s)} \nonumber \\
					&\times& f^s \left[ (E_{a' (N_C + s)} - E_{a N_C} - \mu_r) / k_B T_r \right], \nonumber \\
\end{eqnarray}
and $K^{s}_{a N_C, a' (N_C + s)} = \sum_r K^{r, s}_{a N_C, a' (N_C + s)}$.
Since we do not assume a homogeneous monolayer, the tunnel rates, $\Gamma_{a N_C, a' (N_C + s)}$, depend on both the initial and final 
many-body eigenstates. In particular, $\Gamma_{a N_C, a' (N_C + s)} = 0$ whenever $|a N_C \rangle$ and $|a' (N_C + s)\rangle$ do not 
simply differ by the occupation of a single molecule, since this is all that can be changed by a single SET process.

Equation~(\ref{SME}) simply expresses the fact that the change in occupation of state $| a N_C \rangle$ is determined by the sum of all 
tunnel processes filling $| a N_C \rangle$, weighted by the occupations of the corresponding initial states, minus all tunnel 
processes emptying $| a N_C \rangle$. In the steady state limit, the left hand side of Eq.~(\ref{SME}) is set to zero.
Probability normalization is enforced by Eq.~(\ref{Snormalize}). The current from electrode $r$, Eq.~(\ref{Scurrent}), is given by the sum 
of all processes involving an electron tunneling into the monolayer from electrode $r$, minus the sum of all processes where an electron 
tunnels out of the monolayer into electrode $r$, weighted by the corresponding occupation probabilities.

\section{Comparison of different approximations}\label{sec:appE}
\begin{figure*}[b!]
  \includegraphics[height=0.68\linewidth]{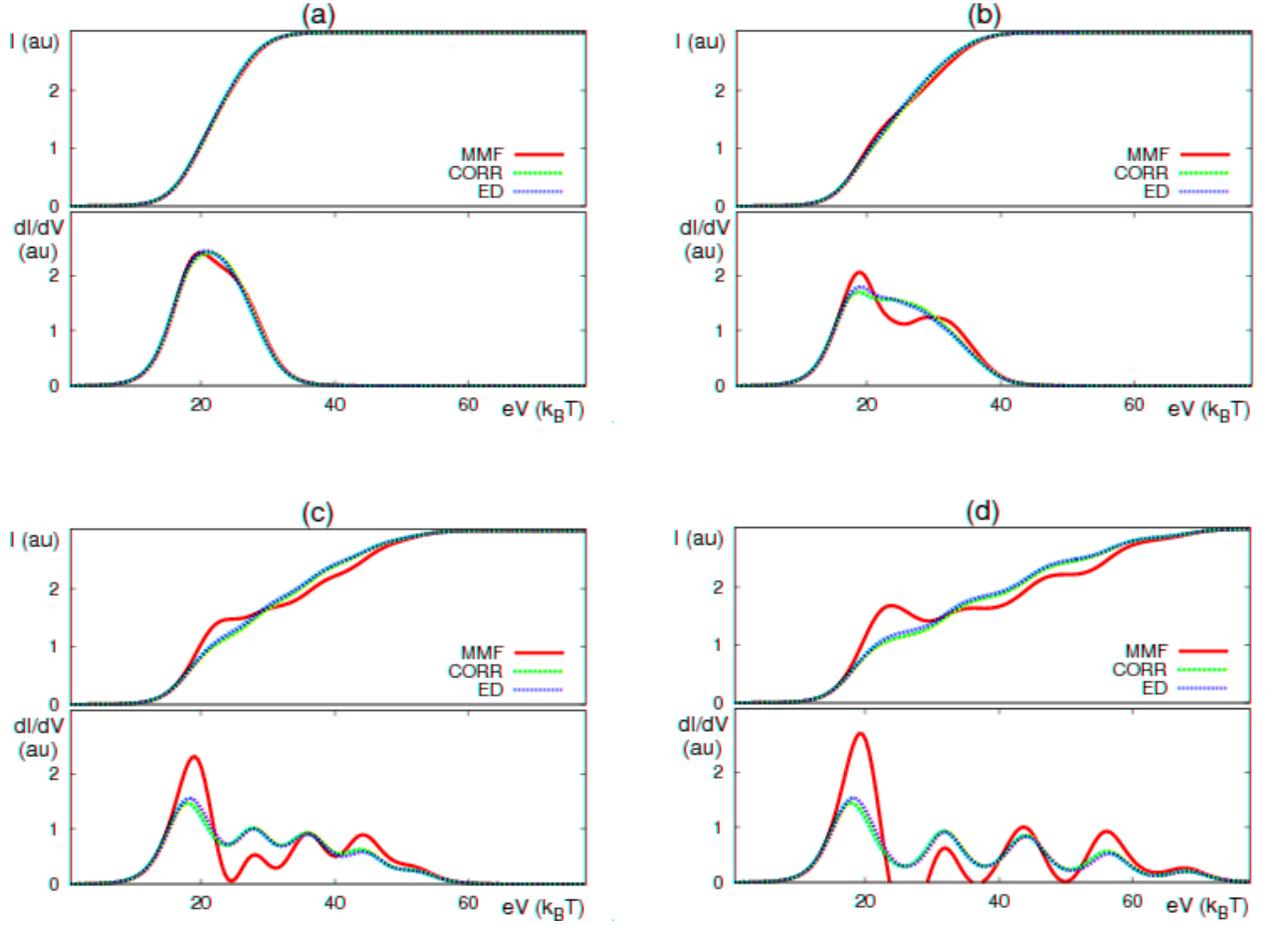}
  \caption{
    \label{fig:S1}
    Current (upper panel) and differential conductance (lower panel) as a function of applied bias voltage. 
    In each subfigure, results using the three different methods described above are shown:
    Mixed mean-field approximation (MMF), truncation of correlations (CORR) and exact diagonalization of a finite monolayer (ED).
    We have everywhere used $\epsilon = 10 k_B T$ and $N = 4$ (square lattice). In ED, we used a square lattice with $3\times3$ molecules, all with the 
    same properties, and periodic boundary conditions [using instead open boundary conditions (not shown) does not qualitatively alter the results, but 
    merely leads to a small increase in the height of the first peak and reduced height of the satellite peaks, reflecting the fact that interactions are somewhat 
    less important in this case]. The strength of the inter-molecular Coulomb interaction is increased, with $U = k_B T$ 
    in (a), $U = 2k_B T$ in (b), $U = 4 k_B T$ in (c), and $U = 6 k_B T$ in (d).}
\end{figure*}
In Fig.~\ref{fig:S1}, we compare the result of the different methods discussed above for different strengths of the inter-molecular 
Coulomb interaction, $U$. For $U = 0$ (not shown), all methods give identical results.
Over the whole range of interaction strengths, we find excellent agreement between the results obtained by 
truncating the correlation, and those from the exact diagonalization of a finite monolayer. 
For $U \gg k_B T$, the mixed mean-field 
solution is clearly inaccurate, predicting the correct height of the high-voltage current plateau, as well as the correct position 
of the current steps, but failing to capture their height, and even 
erroneously predicting NDR for $U = 6k_B T$. Note that this method is not used in any of the results in Figs.~\ref{fig:2}--\ref{fig:4}. 
\bibliographystyle{apsrev}

\end{document}